\magnification=\magstep1
\font\foot=cmr9
\font\summ=cmr9
\font\sum=cmti9
\font\gr=cmbx10 scaled \magstep1
\font\nom=cmcsc10
\font\gt=cmbx12 scaled \magstep2
\centerline{\gt Microscopic Fields and Macroscopic Averages}
\centerline{\gt in Einstein's Unified Field Theory\footnote{*}{\foot Annales de la Fondation Louis de Broglie, Volume 21, 11-38 (1996).}}
\vbox to 0.8 cm {}
\centerline{\nom S. Antoci}\vbox to 0.8 cm {}

\centerline{\summ Dip. di Fisica ``A. Volta'', Via Bassi 6, 27100 Pavia,
Italia}\vbox to 1.8 cm{}

{\summ \noindent ABSTRACT. The problem of the relation between microscopic and macroscopic 
reality in the generally covariant theories is first considered, 
and it is argued that a sensible definition of the macroscopic 
averages imposes a restriction of the allowed transformations of 
coordinates to suitably defined macroscopic transformations. 
Spacetime averages of the geometric objects of a generally 
covariant theory are then defined, and the reconstruction of some 
features of macroscopic reality from hypothetic microscopic 
structures through such averages is attempted in the case of the 
geometric objects of Einstein's unified field theory. It is shown 
with particular examples how a fluctuating microscopic structure 
of the metric field can rule the constitutive relation for 
macroscopic electromagnetism both in vacuo and in nondispersive 
material media. Moreover, if both the metric and the skew field 
${\bf a}^{ik}$ that represents the electric displacement and the 
magnetic field are assumed to possess a wavy microscopic 
behaviour, nonvanishing average generalized force densities 
$<{\bf T}^m_{k;m}>$ are found to occur in the continuum, that originate from a 
resonance process, in which at least three waves need to be 
involved. The previously required limitation of covariance to the 
macroscopic transformations ensures meaning to the notion of a 
periodic microscopic disturbance, for which a wave four-vector can 
be defined. Let $k^A_m$ and $k^B_m$ represent the wave four-vectors of two 
plane wave disturbances displayed by $a^{ik}$, while $k^C_m$ is the wave 
four-vector for a plane wave perturbation of the metric; it is 
found that $<{\bf T}^m_{k;m}>$ can be nonvanishing only if the three-wave 
resonance condition $k^A_m\pm k^B_m\pm k^C_m=0$, so ubiquitous in quantum 
physics, is satisfied. A particular example of resonant process is 
provided, in which $<{\bf T}^m_{k;m}>$ is actually nonvanishing. The wavy 
behaviour of the metric is essential for the occurrence of this 
resonance phenomenon.}\par\vbox to 1.0 cm {}

\noindent {\summ R\'ESUM\'E. }{\sum On examine d'abord le probl\`eme de la relation entre la r\'ealit\'e microscopique et la r\'ealit\'e macroscopique dans les th\'eories covariantes g\'en\'erales, et il est montr\'e qu'une bonne d\'efinition des moyennes macroscopiques impose une restriction aux transformations de coordonn\'ees permises pour le cas macroscopique. On d\'efinit ensuite les moyennes dans l'espace-temps des objects g\'eom\'etriques d'une th\'eorie covariante g\'en\'erale. La reconstitution de certaines propri\'et\'es de la r\'ealit\'e macroscopique 
\`a partir de structures microscopiques suppos\'ees est ensuite tent\'ee dans le cas des objects g\'eom\'etriques de la th\'eorie du champ unifi\'e d'Einstein. Il est montr\'e par des exemples, comment une structure microscopique fluctuante du champ de la m\'etrique peut r\'egir la relation constitutive de l'\'electromagn\'etisme macroscopique, dans le vide comme dans des milieux mat\'eriels non-dispersifs. De plus, si la m\'etrique, et le champ antisym\'etrique ${\bf a}^{ik}$, qui repr\'esente le d\'eplacement \'electrique et le champ magn\'etique, sont suppos\'es avoir un comportement microscopique oscillant, les densit\'es de force g\'en\'eralis\'ees, non-nulles en moyenne, $<{\bf T}^m_{k;m}>$ r\'evel\'ent leur existence dans le milieux continu, et elles viennent d'un processus de r\'esonance impliquant trois ondes. La limite de la covariance, n\'ecessaire pour les transformations macroscopiques, assure une signification \`a la notion de perturbation p\'eriodique microscopique, pour laquelle on peut d\'efinir un quadrivecteur d'onde. On d\'esigne par $k^A_m$ et $k^B_m$ les quadrivecteurs d'onde de deux perturbations sous forme d'ondes planes repr\'esent\'ees par 
$a^{ik}$, tandis que $k^C_m$ repr\'esente le quadrivecteur d'onde pour une onde plane perturbant la m\'etrique; on trouve que 
$<{\bf T}^m_{k;m}>$ ne peut \^etre non-nulle que si la condition de r\'esonance des trois ondes $k^A_m\pm k^B_m\pm k^C_m=0$, si omnipr\'esente en th\'eorie quantique, est satisfaite. Un cas particulier de processus r\'esonant est pr\'esent\'e dans lequel $<{\bf T}^m_{k;m}>$ est effectivement non-nulle. L'existence de ce ph\'enomene de r\'esonance repose essentiellement sur le comportement oscillant de la m\'etrique.}
\par\vbox to 1.8 cm {}

\noindent{\gr 1. INTRODUCTION: SOME REMARKS ON THE RELATION BETWEEN MACRO AND MICROPHYSICS IN A GENERALLY COVARIANT THEORY}\par\vbox to 1.0 cm {}

The issue of the relation between macroscopic and microscopic 
reality as viewed through the evolution of the physical theories 
is a quite complex, curious problem, whose attempted solutions 
seem to reflect more the idiosyncrasies of the inquiring mind than 
an actual structure of the world. Since the direct perception of a 
microscopic reality is per se beyond the capability of our 
unassisted senses, we could have dispensed ourselves altogether 
with developing theories about such a hypothetical entity, had we 
not perceived the existence of certain macroscopic structures or 
processes, whose explanation looked possible through the 
hypothesis of some chain of causes and effects, or of some 
cooperative process, that related the macroscopic phenomenon 
displaying these structures or processes to underlying microscopic 
occurrences; since the regularities of these macroscopic phenomena 
were not dissimilar from the ones present in other situations, 
when no hint of a microscopic substructure was apparent, they also 
seemed amenable to a rational understanding, and we were forced to 
give up the comfortable ideal of a macroscopic physical theory 
closed in itself.\par
The natural philosopher confronted henceforth a difficult trial 
and error game, initially played along the following, somewhat 
circuitous route: he aimed at describing macroscopic reality in 
all its occurrences, but his mind could confidently avail only of 
the concepts and of the laws belonging to that sort of macroscopic 
theories that were just silent about the macroscopic occurrences 
whose explanation appealed to a microscopic substructure. As a 
first move, he then selected among his macroscopic concepts and 
laws the ones that, for some faith in the uniformity and in the 
simplicity of the world, he felt inclined to believe valid also at 
a microscopic scale, and used them for describing the behaviour of 
hypothetical microscopic structures, again imagined as simple, 
idealized replicas of some objects of macroscopic experience. The 
route back to macroscopic reality was then attempted via 
statistical hypotheses and 
methods.\par
Of course, if he fails to produce in this way a better theory of 
macroscopic reality with respect to the ones from which he drew 
inspiration, the natural philosopher can point an accusatory 
finger in several directions, for either the concepts and the laws 
that he has chosen to transfer to a small scale, or the 
microscopic structures that he has imagined, or else the 
statistical methods that he has availed upon may represent or 
include faulty assumptions.\par
An unusual amount of creativity is required at this point in order 
to divine what coordinated changes of the chosen concepts and 
laws, what invention of new microscopic structures, what new 
assumptions about the statistical behaviour, possibly without 
counterpart in the macroscopic experience, may result in a less 
unsatisfactory reconstruction of the macroscopic world. Hopeless 
as it may seem, this approach has led from Newtonian dynamics and 
Maxwell's theory through the electron theory of Lorentz to the 
Planck-Einstein-Bohr theory, then to matrix mechanics, to 
Schr\"odinger's equation, to Dirac's equation and to quantum 
electrodynamics. In retrospect, it is surprising how much 
heuristic value was already contained in the starting point chosen 
by Lorentz [1], how many qualitative features of macroscopic 
reality could already be accounted for by simply transferring to a 
small scale the knowledge gathered about macroscopic dynamics and 
macroscopic electromagnetism in vacuo. The heuristic value of 
Lorentz' attempt did not vanish even after its failure was 
ascertained, but persisted under several respects also through the 
subsequent developments; quantum mechanics and quantum 
electrodynamics may be viewed as the outcome of the efforts aimed 
at understanding what changes in the concepts, in the laws and in 
the statistical assumptions needed to be introduced in order to 
lead to completion Lorentz' program without renouncing two of its 
basic tenets: the adoption of Maxwell's electromagnetism in vacuo 
as a formal ingredient relevant at a microscopic scale, and the 
associated concept of the charged point particle.\par
The whole transition from the electron theory of Lorentz to 
quantum electrodynamics occurred by retaining the inertial 
reference frame as the appropriate spacetime background for the 
physical processes; general relativity, which appeared during that 
transition, had no r\^ole in it: the empirical confirmation achieved 
by general relativity in correcting certain small discrepancies 
between Newton's gravitodynamics and the astronomical observations 
led to view this theory of Einstein in a purely macroscopic 
perspective, and its essential novelties, like the abandonment of 
the inertial frame and its unique interplay between matter and 
spacetime structure appeared, apart from notable exceptions [2,3], 
as useless complications in the difficult task of providing, 
through a careful formulation of hypotheses of a microscopic 
character, a precise account of the manifold aspects of matter and 
of radiation. The problem of the relation between macro and 
microphysics in the generally covariant theories started to 
attract considerable attention only when quantum theory had 
evolved, in the minds of the theoreticians, from the instrumental 
condition of a set of hypotheses well suited to overcome the 
failure of the electrodynamic program initiated by Lorentz to the 
status of a general system of axioms, prescribing the formal 
framework within which any field theory must be inscribed in order 
to become properly ``microscopic'' and hopefully entitled to provide 
a physically correct answer once the way back to the macroscopic 
scale is suitably completed through statistical methods. It was 
then felt that general relativity, as the best available theory of 
the gravitational field, had to be quantized: the principle of 
uniformity in the description of the physical world seemed to 
impose this task.\par
If, confronted with the issue of quantizing general relativity, 
and with the long sequence of conceptual and of technical impasses 
that this attempt has encountered since its inception, we look 
back for inspiration at the path that has led from the original 
program of Lorentz to quantum electrodynamics, we note that each 
new step along that path has been taken in order to overcome some 
defect or failure in the description of the experimental facts by 
the theoretical model achieved in the previous step, while in the 
case of gravitation one notes a disconcerting lack of constraining 
experimental evidence comparable to, say, the existence of the 
Balmer lines or of the blackbody radiation spectrum, that would 
provide guidance and dispel the dangers of academicism from a 
theoretical effort otherwise motivated by essentially formal 
reasons.\par
General relativity, however, is not just a field theory for 
macroscopic gravitation; it looks rather like the first, 
provisional achievement of a program aimed at representing the 
whole of physical reality in a new way that dispenses with the 
need of the inertial reference frame and posits a direct relation 
between spacetime structure and material properties; due to these 
essential novelties, common to all the generally covariant 
theories, one should be prepared to acknowledge that for these 
theories the issue of the relation between macro and microphysics 
may well require a totally different approach from the one 
successfully adopted with the theories that retain the inertial 
frame; it may be more appropriate then to draw free inspiration 
from the historical sequence of attempts that has led from the 
electron theory by Lorentz to quantum mechanics, rather than stick 
to the formal expression of the end results of that endeavour, 
that was rooted in so different a conceptual framework.\par
According to this spirit one could try, as a first attempt, to 
transfer the driving ideas of Lorentz' program in the new 
environment, i.e. one should select concepts and laws from the 
available generally covariant theories and tentatively extrapolate 
them to a small scale; one should then invent microscopic 
structures built up with the geometric objects [4] of these 
theories and try the way back to macroscopic reality via 
statistical assumptions and methods. Due to the nonlinearity of 
the generally covariant theories, totally new possibilities will 
appear along the back and forth route between micro and 
macrophysics, as it was already intimated, in the framework of the 
Riemannian geometry, by the investigations performed by C. Lanczos 
[5,6]. These new possibilities are by no means confined to the 
realm of gravitational physics; we shall not fear the risk of 
academicism, since the whole of the experimental knowledge 
gathered about the structure and the behaviour of matter and of 
radiation will be in principle at our disposal for testing the 
validity of concepts, structures and laws that we may 
propose.\vfill\break

\noindent{\gr 2. MACROSCOPIC AVERAGES IN THE GENERALLY COVARIANT 
THEORIES}\par\vbox to 0.8 cm {}

In a generally covariant framework the very definition of 
macroscopic averages deserves a close scrutiny, as it is intimated 
e.g. by the investigations [7] dealing with such averages in 
cosmology. Here we wish to point out that the definition of a 
macroscopic average appears related to general covariance in a 
peculiar way. Imagine that one chooses a given generally covariant 
theory and tentatively assumes that its geometric objects and its 
laws are meaningful at any scale; their mathematical expression 
permits this hypothesis. The assumed general covariance of the 
theory will allow for general transformations of coordinates 
$x'^i=f^i(x^k)$ in which the functions $f^i$ need only to satisfy 
appropriate conditions of continuity, differentiability, and 
regularity of the functional determinant 
$det(\partial x'^i/{\partial x^k})$, 
but are otherwise arbitrary. For instance 
$f^i(x^k)$ can display a microscopic structure; this possibility is 
consistent with the behaviour of a geometric object 
$O^{mn..}_{ik..}(x^p)$ at a 
microscopic scale. In the following we shall write simply $O(x^p)$ 
for the generic geometric object whenever this shorthand does not 
cause confusion. Assume now that the way back from micro to 
macrophysics entails some averaging process, performed in the 
coordinate system $x^i$, through which some average quantity{\footnote
{\foot$^1$}{Unlike the expression ``geometric object'', here and in the 
following, the word ``quantity'' is not used in the strict technical sense of Ref. 4.}} 
$\bar O(x^p)$ 
is extracted from the unaveraged one $O(x^p)$ by some mathematical 
procedure, intended to mimic a process of measurement performed 
through a macroscopic device in the reference frame associated 
with the coordinate system $x^i$. In compliance with our ideas about 
averages and macroscopic reality, we expect that in the average 
field $\bar O(x^p)$ all the microscopic structures displayed by 
$O(x^p)$ are completely effaced, i.e., that we can associate to a 
generic point $x^i_0$ a box containing the points for which

$$\vert x^i-x^i_0\vert<\beta^i,\eqno(1)$$
where the positive numbers $\beta^i$ are exceedingly large with 
respect to the increments $\vert x^i-x^i_0\vert$ over which a variation 
of $O(x^p)$ becomes perceptible, and yet exceedingly small with 
respect to the increments $\vert x^i-x^i_0\vert$ over which a variation 
of $\bar O(x^p)$ can be appreciated.\par
In what manner shall $\bar O(x^p)$ behave under a coordinate 
transformation? It seems desirable that the quantity $\bar O(x^p)$ 
replicate the transformation properties of $O(x^p)$ but, if we 
insist that the averaged field shall transform like the unaveraged 
one under all the admissible transformations of coordinates [8], 
our expectation about the effacement of the microscopic structures 
cannot be realized. In fact, let us assume for instance that the 
components of $\bar O(x^p)$ display a complete cancellation of the 
microscopic structure; a transformation of coordinates $x^i=f^i(x^k)$ 
exhibiting some microscopic vagary will suffice in reintroducing 
the unwanted microscopic structure in the components 
$\bar O'^{mn..}_{ik..}(x^p)$  
of the average field, defined with respect to the primed 
coordinate system.\par
The effacing ability of the averaging procedure is compatible with 
the requirement that $\bar O(x^p)$  transform according to the same 
rule obeyed by the geometric object $O(x^p)$  only for the subset of 
transformations $x'^i=h^i(x^k)$ such that, if $h^i$ and its derivatives up 
to some appropriate order are expanded in Taylor's series around 
the generic point $x^i_0$:

$$x'^i=h^i(x^k_0)+(h^i_{,m})_0(x^m-x^m_0)
+(1/2)(h^i_{,m,n})_0(x^m-x^m_0)(x^n-x^n_0)+...,\eqno(2)$$

$$x'^i_{,m}=(h^i_{,m})_0+(h^i_{,m,n})_0(x^n-x^n_0)+...,\eqno(3)$$
and so on, the leading term in each expansion is exceedingly 
larger than the subsequent ones for all the points $x^i$ within the 
box defined by (1). In order to retain the distinction between a 
microscopic and a macroscopic scale we shall admit only coordinate 
transformations that fulfil this condition{\footnote
{\foot$^2$}{The idea that a restriction of covariance is needed in order to establish a distinction between a macroscopic and a microscopic scale is present in the literature that deals with the problem of defining averages in cosmology. See e.g. the Introduction of the paper by A. H. Nelson (Ref. 7), where such a restriction is invoked as a necessary means of discriminating between the global and the local properties of the metric, and the paper by T. W. Noonan (Ref. 9). According to the latter author we must postulate a duality, i.e. the existence of two types of observers, a macroscopic observer who can see only the large-scale properties of the medium, and a microscopic observer who can see the small-scale properties. As regards the allowed coordinate transformations, it is the macroscopic observer who, according to Noonan, imposes the greater constraint, since he is by definition unable to perceive coordinate transformations endowed with a microscopic structure.}}; they will be called 
henceforth macroscopic, and also macroscopic will be called all 
the coordinate systems that can be reached from the coordinate 
system $x^i$ through a macroscopic transformation, provided that the 
average quantities defined in the system $x^i$ display the required 
effacement of the microscopic structures.\par
Up to now no word has been said about the definition of average 
that we intend to adopt; the necessary restriction of covariance 
to the previously defined macroscopic transformations eases the 
problem, since the very form in which this restriction was 
expressed suggests the following simple scheme of spacetime 
averaging [9] as an admissible choice. Let us begin by introducing 
in the spacetime region where the averages must be defined a 
coordinate system $x^i$, and by associating to each point $x^i$ a 
neighbourhood $\Omega(x^i)$ according to the following prescription: 
we surround a given point $x^i_0$ with a box defined by (1), that we 
choose as $\Omega(x^i_0)$; the neighbourhood $\Omega(x^i_0+\delta x^i)$ 
associated with the point whose coordinates are $x^i_0+\delta x^i$ is 
then simply defined as the box for which

$$\vert x^i-x^i_0-\delta x^i\vert<\beta^i;\eqno(4)$$
in this way a neighbourhood is associated to each point in the 
spacetime region under question. This association will be kept in 
all the allowed coordinate systems, i.e. if $x'^i$ and $x^i$ denote the 
same point in two coordinate systems, $\Omega(x'^i)$ shall contain 
the same points as $\Omega(x^i)$. The spacetime average of the field 
$O(x^p)$ is then defined as

$${\bar O}(x^p)\equiv<O>_{\Omega(x^p)}
={{\int_{\Omega(x^p)}Od\Omega}
\over{\int_{\Omega(x^p)}d\Omega}};\eqno(5)$$
it is a field that, with adequate accuracy, transforms under the 
macroscopic coordinate transformations according to the same law 
as the geometric object $O(x^p)$ ; moreover, in the coordinate 
system $x^i$ one finds exactly

$$<O_{,r}>_{\Omega(x^p)}={\bar O}_{,r}(x^p),\eqno(6)$$
and the same property holds with adequate accuracy in all the 
macroscopic coordinate systems.\par\vbox to 1.2 cm{}

\noindent{\gr 3. THE GEOMETRIC OBJECTS OF THE UNIFIED FIELD THEORIES OF EINSTEIN 
AND SCHR\"ODINGER}\par\vbox to 0.8 cm {}

We have stressed that the nonlinearity of the generally covariant 
theories offers completely new possibilities in the back and forth 
game of reconstructing the macroscopic reality from hypothetical 
microscopic structures and laws; this paper aims at evidencing two 
such possibilities offered through the geometric objects of the 
non-Riemannian theories developed by Einstein and by Schr\"odinger 
in their search [10,11] for an extension of the general relativity 
of 1915 that could encompass both gravitation and 
electromagnetism. In retrospect, one does not see really cogent 
reasons why these theories should provide, as hoped for by their 
authors, field theoretical completions of general relativity: 
their geometric objects are so closely akin to the ones occurring 
in that theory, that one may well wonder [12] why in such theories 
one should depart from the attitude kept in general relativity, 
where one has not to do with field laws describing the evolution 
of matter, but rather with a fundamental definition of the 
stress-momentum-energy tensor in terms of the metric [13,14]. It 
seems reasonable to assume that a similar situation should prevail 
also in the above mentioned field theories, and to investigate 
what new definitions of physical quantities can be given through 
the geometric objects first envisaged by Einstein.\par
One possible identification [15] of those geometric objects with 
physical entities runs as follows: in a four-dimensional manifold 
endowed with real coordinates $x^i$ a nonsymmetric tensor density 
${\bf g}^{ik}$ defines [16] through its symmetric part ${\bf g}^{(ik)}$ 
the metric tensor $s_{ik}$:

$${\bf s}^{ik}={\bf g}^{(ik)},~~~{\bf s}^{ik}=(-s)^{1/2}s^{ik},~~~
s^{im}s_{km}=\delta^i_k,~~~s=det(s_{ik}),\eqno(7)$$
while its skew part ${\bf g}^{[ik]}\equiv{\bf a}^{ik}$ defines the 
electric displacement $\bf D$ and the magnetic field $\bf H$ through 
the identifications:

$$({\bf a}^{41},{\bf a}^{42},{\bf a}^{43})\Rightarrow(D_1,D_2,D_3),~~~
({\bf a}^{23},{\bf a}^{31},{\bf a}^{12})
\Rightarrow(H_1,H_2,H_3);\eqno(8)$$
the electric four-current density ${\bf j}^i$ is correspondingly 
defined as

$${\bf j}^i=(1/{4\pi}){\bf g}^{[is]}_{~~~~,s}.\eqno(9)$$
Through the equation

$${\bf g}^{qr}_{~~,p}+{\bf g}^{sr}\Gamma^q_{sp}+{\bf g}^{qs}\Gamma^r_{ps}
-{\bf g}^{qr}\Gamma^t_{(pt)}
=(4\pi/3)({\bf j}^q\delta^r_p-{\bf j}^r\delta^q_p)\eqno(10)$$
the tensor density ${\bf g}^{ik}$ uniquely [17,18] determines the 
nonsymmetric affine connection $\Gamma^i_{km}$, by definition constrained 
to yield $\Gamma^k_{[ik]}=0$, through which the symmetrized Ricci 
tensor

$$B_{ik}(\Gamma)=\Gamma^a_{ik,a}
-(1/2)(\Gamma^a_{ia,k}+\Gamma^a_{ka,i})
-\Gamma^a_{ib}\Gamma^b_{ak}+\Gamma^a_{ik}\Gamma^b_{ab}\eqno(11)$$
is constructed [19]. The reason why this symmetrized tensor is 
considered in place of the plain one occurring in general 
relativity will soon be apparent. The symmetric part of $B_{ik}$ is 
assumed to define the symmetric stress-momentum-energy tensor $T_{ik}$ of 
a material medium through the equation

$$B_{(ik)}(\Gamma)=8\pi(T_{ik}-(1/2)s_{ik}s^{pq}T_{pq}),\eqno(12)$$
while its skew part $B_{[ik]}$ is identified with the electric field 
$\bf E$ and with the magnetic induction $\bf B$ through the rule

$$(B_{[14]},B_{[24]},B_{[34]})\Rightarrow(E_1,E_2,E_3),~~~
(B_{[23]},B_{[31]},B_{[12]})\Rightarrow(B_1,B_2,B_3);\eqno(13)$$
the magnetic four-current $K_{ikm}$ is consequently defined as

$$K_{ikm}=(3/{8\pi})B_{[[ik],m]},\eqno(14)$$
where $B_{[[ik],m]}\equiv(1/3)(B_{[ik],m}+B_{[km],i}+B_{[mi],k})$. Thanks to equation (10) and to the definition (11) it happens that, if $T_{ik}$, ${\bf j}^i$ and $K_{ikm}$ are 
the material counterpart of a given field ${\bf g}^{ik}$, the matter 
counterpart of the transposed field ${\bf \tilde g}^{ik}\equiv
{\bf g}^{ki}$, that we indicate with $\tilde T_{ik}$, ${\bf \tilde j}^i$ 
and $\tilde K_{ikm}$, is 
such that $\tilde T_{ik}=T_{ik}$, ${\bf \tilde j}^i=-{\bf j}^i$ and 
$\tilde K_{ikm}=-K_{ikm}$, i.e. ``the requirement that positive and negative 
electricity enter symmetrically into the laws of physics'' [20] is 
satisfied. When ${\bf j}^i$ is not vanishing, this requirement 
cannot be fulfilled if, instead of $B_{ik}$, the plain Ricci tensor is 
adopted.\par
The consistency of the identifications introduced above appears 
from the contracted Bianchi identities, that can be written [15] 
as

$${\bf T}^m_{k;m}
=(1/2)({\bf j}^iB_{[ki]}+K_{ikm}{\bf g}^{[mi]}),\eqno(15)$$
where ${\bf T}^m_k={\bf s}^{mi}T_{ki}$, and ``;'' indicates the 
covariant differentiation performed with the Christoffel affine 
connection

$$\Sigma^i_{km}=(1/2)s^{ia}(s_{ak,m}+s_{am,k}-s_{km,a})\eqno(16)$$
associated with the metric $s_{ik}$ (that will be hereafter used to 
move indices, to build tensor densities from tensors, and 
vice-versa). From (15) one gathers than the local nonconservation 
of the energy tensor in the Riemannian spacetime defined by the 
metric $s_{ik}$ is due to the Lorentz coupling of the electric 
four-current to $B_{[ik]}$ and of the magnetic four-current to 
${\bf g}^{[ik]}$, 
as one expects to occur in the electrified material medium of a 
gravito-electromagnetic theory. Two versions of this theoretical 
structure are possible, according to whether ${\bf g}^{ik}$ is a real 
nonsymmetric, or a complex Hermitian tensor density.\vfill\break

\noindent{\gr 4. THE CONSTITUTIVE RELATION FOR MICROSCOPIC 
ELECTROMAGNETISM}\par\vbox to 0.8 cm {}

Assuming, as we are doing, that the sort of electromagnetism that 
we are reading off the geometric objects of Einstein's unified 
field theory is competent at a microscopic scale means a 
substantial departure from the letter of Lorentz' approach. In 
that case, a simple hypothesis is made for the relation between 
inductions and fields that should prevail microscopically: if the 
skew tensor density ${\bf a}^{ik}$ represents as before the electric 
displacement and the magnetic field, the skew tensor $b_{ik}$ that 
defines the electric field and the magnetic induction is given by

$$b_{ik}=a_{ik}\equiv(-s)^{-1/2}s_{ip}s_{kq}{\bf a}^{pq},\eqno(17)$$
i.e. by an algebraic expression in terms of $s_{ik}$ and of 
${\bf a}^{ik}$, 
in which ${\bf a}^{ik}$ enters in a linear way. We have written this 
constitutive relation in curvilinear coordinates for contrasting 
it with the one that exists instead between ${\bf a}^{ik}$ and 
$B_{[ik]}(\Gamma)$, a nonlinear, differential relation which is the 
antisymmetric counterpart of the relation between the metric 
tensor density ${\bf s}^{ik}$ and the symmetric field $B_{(ik)}$ that 
defines 
through (12) the stress-momentum-energy content of the manifold; 
both these relations are simultaneously found by solving (10) for 
$\Gamma^i_{km}$ and by substituting its expression in (11). Let

$$S^i_{kmn}=\Sigma^i_{km,n}-\Sigma^i_{kn,m}
-\Sigma^i_{am}\Sigma^a_{kn}+\Sigma^i_{an}\Sigma^a_{km}\eqno(18)$$
be the Riemann tensor defined with the Christoffel symbol 
$\Sigma^i_{km}$, and assume that ${\bf a}^{ik}$ is a vanishingly small 
quantity; the linear approximation to $B_{[ik]}$ then reads [15]

$$B_{[ik]}=(2\pi/3)(j_{i,k}-j_{k,i})
+(1/2)(a^{~n}_iS_{nk}-a^{~n}_kS_{ni}+a^{pq}S_{pqik}
+a^{~~;a}_{ik;a}),\eqno(19)$$
where $S_{ik}\equiv S^p_{ikp}$ is the Ricci tensor of $s_{ik}$ and $a^{~~;m}_{ik}\equiv s^{mn}a_{ik;n}$. This 
equation shows how widely the constitutive relation for 
microscopic electromagnetism that we are adopting departs from the 
one assumed by Lorentz already for weak inductions and fields. The 
right-hand side of (19) is homogeneous of degree two with respect 
to differentiation; therefore the small scale behaviour of both 
${\bf a}^{ik}$ and ${\bf s}^{ik}$ will be crucial in ruling the relation 
between inductions and fields, as it is fundamental in determining 
the stress-momentum-energy content of matter; the same assertion 
holds for the generalized force density felt by the electrified 
medium, given by (15). As a consequence, a whole new range of 
possibilities is offered in the game of reconstructing macroscopic 
reality from a hypothetic microscopic behaviour, which has no 
counterpart in theories in which the constitutive relation (17) is 
instead adopted at a microscopic scale.\vfill\break

\noindent{\gr 5. MICROSCOPIC FLUCTUATIONS OF THE METRIC CAN RULE THE MACROSCOPIC CONSTITUTIVE RELATION IN VACUO AND IN NONDISPERSIVE MEDIA}\par\vbox to 0.8 cm {}

Imagine for instance that, while ${\bf a}^{ik}$ is very small and 
varying only at a macroscopic scale, $s_{ik}$ exhibits a microscopic 
structure. We can write

$$s_{ik}={\bar s}_{ik}+\delta s_{ik},\eqno(20)$$
where ${\bar s}_{ik}$ means the average metric calculated according to the 
definition (5), while $\delta s_{ik}$ indicates a microscopic fluctuation. We shall assume that $\vert \delta s_{ik}\vert$ is very small with respect 
to $\vert s_{ik}\vert$, and that 
$\vert \delta s_{ik}\vert \ll\vert \delta s_{ik,m}\vert \ll
\vert \delta s_{ik,m,n}\vert$, since the characteristic length 
of the fluctuations is microscopic, i.e. quite small in our units. 
What is the behaviour of the average field ${\bar B}_{[ik]}$ under these 
conditions? We can avail of the expression (19) in order to 
provide a first answer, limited to the linear approximation in 
${\bf a}^{ik}$. This expression can be expanded into a sum of addenda, 
each one given by ${\bf a}^{ik}$, or ${\bf a}^{ik}_{~~,m}$, or else 
${\bf a}^{ik}_{~~,m,n}$, times a product of several terms, individually given by $s_{ik}$, $s^{ik}$, 
$(-s)^{-1/2}$ and by the ordinary derivatives of $s_{ik}$ up to second order, that we call metric factor, because only the metric appears in it; 
in a metric factor containing $s_{ik,m,n}$ no further derivatives are 
allowed. Since ${\bf a}^{ik}$ is assumed to vary at a macroscopic 
scale, averaging the individual addendum reduces to calculating 
the mean of the corresponding metric factor. Let us turn each 
metric factor displaying a second derivative into the overall 
derivative of a metric factor in which $s_{ik}$ is differentiated once, 
minus the sum of metric factors that contain the product of two 
first derivatives. Due to the previously made assumptions and to 
(6), the whole problem of averaging $B_{[ik]}$ thus reduces to evaluating 
the means of metric factors where only the metric and its first 
derivatives are present; the latter can appear at most twice in a 
given metric factor.\par
The mean of a metric factor where no derivatives appear is known, 
since , due to the smallness of the fluctuations, we can write

$$<s_{ik}..s^{pq}..(-s)^{-1/2}>
=\bar s_{ik}..\bar s^{pq}..(-\bar s)^{-1/2};\eqno(21)$$
we assume that the smallness of the fluctuations is so related to 
the shortness of their characteristic length that we can write 
also

$$<s_{ik}..s^{pq}s_{rs,t}..(-s)^{-1/2}>
=\bar s_{ik}..\bar s^{pq}\bar s_{rs,t}..(-\bar s)^{-1/2};\eqno(22)$$
while the evaluation of

$$<s_{ik}..s^{pq}s_{rs,t}s_{uv,z}..(-s)^{-1/2}>
=\bar s_{ik}..\bar s^{pq}<s_{rs,t}s_{uv,z}>
..(-\bar s)^{-1/2};\eqno(23)$$
will require hypotheses of a statistical character on the 
microscopic behaviour of the metric, since $<s_{ik,m}s_{np,q}>$ will 
differ strongly from $\bar s_{ik,m}\bar s_{np,q}$. The quantity

$$F_{ikmnpq}=<s_{ik,m}s_{np,q}>-\bar s_{ik,m}\bar s_{np,q},\eqno(24)$$
which, due to the previous assumptions, behaves as a tensor under 
macroscopic coordinate transformations, is the appropriate object 
for encoding the statistical information required for the explicit 
calculation of ${\bar B}_{[ik]}$.\par
In the particular case, when the fluctuating metric is conformally 
related [21] to its average, i.e. when $s_{ik}=e^\sigma\bar s_{ik}$, 
with $\vert \sigma\vert\ll 1$, we get

$$F_{ikmnpq}=\bar s_{ik}\bar s_{np}<e^{2\sigma}
\sigma_{,m}\sigma_{,q}>=\bar s_{ik}\bar s_{np}c_{mq},\eqno(25)$$
and the statistical information is expressed by the symmetric 
quantity $c_{ik}$, that behaves as a tensor under macroscopic 
transformations. A calculation of ${\bar B}_{[ik]}$ under these 
conditions [22] leads to the result

$$<B_{[ik]}(s_{ab},a_{ab})>
=B_{[ik]}({\bar s}_{ab},{\bar a}_{ab})+D{\bar a}_{ik},\eqno(26)$$
where $D=-(3/2){\bar s}^{pq}c_{pq}$, and the function $B_{[ik]}(s_{ab},a_{ab})$ is given by (19). The first term at the right-hand side of (26) displays on 
the average fields $\bar s_{ik}$ and $\bar a_{ik}$ the same 
dependence that the linear approximation to $B_{[ik]}$ has on 
$s_{ik}$ and $a_{ik}$. The second 
term is just given by the average of $a_{ik}$ times a factor $D$ that 
behaves as a scalar under macroscopic transformations. If $D$ is 
constant in a given spacetime region and its magnitude is such 
that $D\bar a_{ik}$ is by far the dominant term at the right-hand side 
of (26), the usual constitutive relation (17) appropriate to the 
macroscopic vacuum is found to prevail between ${\bf a}^{ik}$ and 
$\bar B_{[ik]}$. Under these circumstances, if the mean magnetic current $\bar K_{ikm}$ is vanishing, as one assumes in macroscopic electromagnetism, one 
finds

$$<B_{[[ik],m]}>=D\bar a_{[[ik],m]}=0,\eqno(27)$$
i.e. the macroscopic inductions and fields fulfil the usual 
equations for vacuum, and the average of the right-hand side of 
(15) exhibits the usual force density felt in vacuo by a 
macroscopic electric four-current ${\bf j}^i$.\par
Although this medium has a weak-field electromagnetic behaviour 
that may exactly reproduce the one appropriate to the macroscopic 
vacuum, its material content is by no means vanishing, not either 
in the average sense, also when $\bar s_{ik}$ is a vacuum metric. Let us 
calculate the mean of the stress-momentum-energy density 
${\bf T}^m_k$; since ${\bf a}^{ik}$ is vanishingly small, we can 
neglect its contribution, and write:

$${\bf T}^m_k={\bf T}^m_k(s_{ab})
={\bf s}^{im}[S_{ik}(s_{ab})-(1/2)s_{ik}S(s_{ab})],\eqno(28)$$
where $S=s^{pq}S_{pq}$. The previous assumptions about the 
fluctuations of $s_{ik}$ suffice also for calculating this 
average [22]; one finds

$$8\pi<{\bf T}^m_k>=8\pi{\bf T}^m_k(\bar s_{ab})
+(3/2)(-\bar s)^{1/2}[\bar s^{im}c_{ik}
-(1/2)\delta^m_k\bar s^{pq}c_{pq}];\eqno(29)$$
therefore the contribution to $<{\bf T}^m_k>$ coming from the 
conformal fluctuations cannot be made to vanish unless $c_{ik}=0$.\par
Fluctuations of the metric with a lesser degree of symmetry can be 
used to mimic the macroscopic constitutive relation in material 
nonconducting media. Let us consider a simple example: suppose 
that a macroscopic coordinate system exists, in which

$$s_{i4}=\bar s_{i4},~~~s_{\lambda\mu}
=e^\sigma\bar s_{\lambda\mu},~~~\vert \sigma \vert \ll 1,\eqno(30)$$
i.e. the spatial components of the metric $s_{\lambda\mu}$ perform conformal fluctuations of very small amplitude and with a very small 
characteristic length around their average, while the other 
components are smooth; Greek indices label the spatial 
coordinates. We assume that besides (21) also (22) still holds; 
due to the choice (30), the nonvanishing components of $F_{ikmnpq}$ 
will be

$$F_{\alpha\beta m\gamma\delta n}=\bar s_{\alpha\beta}
\bar s_{\gamma\delta}<e^{2\sigma}\sigma_{,m}\sigma_{,n}>
=\bar s_{\alpha\beta}\bar s_{\gamma\delta}c_{mn},\eqno(31)$$
and the mean components of $B_{[ik]}$ in the linear 
approximation (19) read

$$\eqalign{<B_{[\lambda\mu]}(s_{ab},a_{ab})>
&=B_{[\lambda\mu]}({\bar s}_{ab},{\bar a}_{ab})\cr
&+(1/8)[{\bar a}^{~\epsilon}_\lambda c_{\mu\epsilon}
-{\bar a}^{~\epsilon}_\mu c_{\lambda\epsilon}
-{\bar a}^{~4}_{\lambda}c_{\mu 4}
+{\bar a}^{~4}_{\mu}c_{\lambda 4}
-5{\bar a}_{\lambda\mu}{\bar s}^{pq}c_{pq}],\cr
<B_{[4\mu]}(s_{ab},a_{ab})>
&=B_{[4\mu]}({\bar s}_{ab},{\bar a}_{ab})\cr
&+(1/8)[{\bar a}^{~\epsilon}_\mu c_{4\epsilon}
-3{\bar a}^{~\epsilon}_4 c_{\mu\epsilon}
-{\bar a}_{4\mu}(9{\bar s}^{\gamma\delta}c_{\gamma\delta}
+12{\bar s}^{44}c_{44})].\cr }
\eqno \hbox{$(32)$} $$
If the first terms at the right-hand sides are negligible with 
respect to the remaining ones, (32) expresses the constitutive 
relation for macroscopic electromagnetism in a linear, 
nondissipative, nondispersive medium, which is spatially 
anisotropic, nonreciprocal{\footnote{\foot$^3$}{Let ${\bf a}^{ik}$ and $b_{ik}$ have the geometric and physical meaning that was attributed to them at the beginning of Section 4. In a linear nondissipative, nondispersive medium they are related by the equation 
${\bf a}^{ik}=(1/2){\bf X}^{ikpq}b_{pq}$, where ${\bf X}^{ikpq}$ 
is the constitutive tensor density of the medium. Let 
$X^{ikpq}\equiv (-s)^{-1/2}{\bf X}^{ikpq}$ 
be the corresponding tensor: a medium is called 
reciprocal if $X^{ikpq}$ is invariant under reversal of the time coordinate; if not, the medium is called nonreciprocal.}} and nonuniform [23], unless more 
specialized assumptions are made for the behaviour of $s_{ik}$; for 
instance if, in the chosen coordinate system, we have

$$\bar s_{ik}=\eta_{ik}\equiv diag(-1,-1,-1,1),~~~
c_{\lambda\mu}=\alpha\eta_{\lambda\mu},~~~
c_{\lambda 4}=0,~~~c_{44}=\beta,\eqno(33)$$
where $\alpha$ and $\beta$ are constants, (32) becomes

$$\eqalign{<B_{[\lambda\mu]}(s_{ab},a_{ab})>
&=B_{[\lambda\mu]}({\bar s}_{ab},{\bar a}_{ab})
-(1/8)(13\alpha+5\beta)\bar a_{\lambda\mu},\cr
<B_{[4\mu]}(s_{ab},a_{ab})>
&=B_{[4\mu]}({\bar s}_{ab},{\bar a}_{ab})
-(1/8)(30\alpha+12\beta)\bar a_{4\mu},\cr }
\eqno \hbox{$(34)$} $$
and, if the first terms at the right-hand sides are negligible 
with respect to the other ones, the electromagnetic medium will be 
uniform, isotropic and reciprocal.\par
     When the contribution of ${\bf a}^{ik}$ to ${\bf T}^m_k$ is neglected, 
the average components of the stress-momentum-energy density of 
the anisotropic, nonreciprocal, nonuniform electromagnetic medium 
read:

$$\eqalign{8\pi<{\bf T}^\nu_\mu(s_{ab})>
&=8\pi{\bf T}^\nu_\mu(\bar s_{ab})
+(1/2)\bar s^{\lambda\mu}c_{\lambda\mu}
-(1/4)\delta^\nu_\mu(\bar s^{\alpha\beta}c_{\alpha\beta}
+3\bar s^{44}c_{44}),\cr
8\pi<{\bf T}^4_\mu(s_{ab})>
&=8\pi{\bf T}^4_\mu(\bar s_{ab})
+(3/2)\bar s^{44}c_{4\mu},\cr
8\pi<{\bf T}^\mu_4(s_{ab})>
&=8\pi{\bf T}^\mu_4(\bar s_{ab})
+(1/2)\bar s^{\mu\lambda}c_{4\lambda},\cr
8\pi<{\bf T}^4_4(s_{ab})>
&=8\pi{\bf T}^4_4(\bar s_{ab})
-(1/4)\bar s^{\alpha\beta}c_{\alpha\beta}
+(3/4)\bar s^{44}c_{44},\cr }
\eqno \hbox{$(35)$} $$
while the nonvanishing components of $<{\bf T}^m_k>$ for the uniform, 
isotropic, reciprocal specialization defined by (33) are

$$\eqalign{8\pi<{\bf T}^\nu_\mu>
&=-(1/4)\delta^\nu_\mu(\alpha+3\beta),\cr
8\pi<{\bf T}^4_4>
&=(3/4)(\beta-\alpha);\cr }
\eqno \hbox{$(36)$} $$
they correspond to a uniform mechanical continuum, endowed only 
with energy density and with an isotropic pressure.
To sum up the results of this Section, we have shown through 
particular examples how microscopic fluctuations of the metric can 
produce dynamically the constitutive relation for weak inductions 
and fields that prevails macroscopically both in vacuo and in 
material nondispersive media, although the microscopic relation 
(19) has a completely different character. These fluctuations 
produce also an average stress-momentum-energy content of the 
continuum, which is however ineffective in ruling the macroscopic 
geometry of spacetime: ${\bf T}^m_k(s_{ab})$ and its average can have quite 
large components despite the fact that $\bar s_{ik}$ is for instance 
everywhere Minkowskian; therefore we find no objection at present 
against the supposed existence of this stress-momentum-energy 
content of the continuum, and of the microscopic behaviour of $s_{ik}$ 
from which it finds its origin.\par\vbox to 1.2 cm{}

\noindent{\gr 6. RESONANCES BETWEEN MICROSCOPIC WAVES OF ${\bf g}^{ik}$ CAN PRODUCE NET AVERAGE GENERALIZED FORCES IN THE MEDIUM}
\par\vbox to 0.8 cm {}

Suppose now that both $s_{ik}$ and ${\bf a}^{ik}$ are endowed with a 
microscopic structure; an intriguing relation appears then between 
a coherent behaviour of the two fields at a microscopic scale and 
the macroscopic generalized forces that show up in the continuum. 
Let us assume for instance that, within a box $\Omega$ defined by 
(1) and with respect to the coordinate system $x^i$, $s^{ik}$ and 
$a^{ik}\equiv(-s)^{-1/2}{\bf a}^{ik}$ can be written as

$$\eqalign{s^{ik}
&=\eta^{ik}+b^{ik}_A sin(k^A_m x^m+\varphi^A),\cr
a^{ik}
&=c^{ik}_A sin(k^A_m x^m+\varphi^A),\cr }
\eqno \hbox{$(37)$} $$
where $\eta^{ik}$ is the Minkowski metric, while $b^{ik}_A=b^{ki}_A$ and 
$c^{ik}_A=-c^{ki}_A$ have 
constant values and are so small that can be dealt with as first 
order infinitesimal quantities; the usual summation rule is 
extended to the upper case Latin index $A=1,..,n$ numbering the 
progressive sinusoidal waves that have $k^A_m$ as wave four-vector and 
$\varphi^A$ as phase constant. When terms not linear in $b^{ik}_A$ can be 
neglected we can write

$$s_{ik}=\eta_{ik}-b_{Aik} sin(k^A_m x^m+\varphi^A),\eqno(38)$$
where the indices in the small quantities $b^{ik}_A$ have been lowered 
with $\eta_{ik}$. We consider waves whose wavelengths and whose periods 
are exceedingly small with respect to the dimensions of the box; 
the restriction of covariance to the macroscopic transformations, 
that was found necessary for obtaining sensible macroscopic 
averages, ensures now that the concept of a microscopic periodic 
disturbance endowed with a wave four-vector and with a phase 
constant is a meaningful one in $\Omega$: the trigonometric 
behaviour of $s^{ik}$ and of $a^{ik}$ defined by (37), that is destroyed in 
general by an arbitrary transformation of coordinates, is in fact 
preserved within the box $\Omega$ by a macroscopic 
transformation.\par
We are looking after the generalized forces that may appear at a 
macroscopic scale in the continuum due to the microscopic 
behaviour of ${\bf g}^{ik}$; the spacetime average

$$<{\bf T}^m_{k;m}>={{\int_{\Omega}{\bf T}^m_{k;m}d\Omega}
\over{\int_{\Omega}d\Omega}}\eqno(39)$$
of the generalized force density over the box $\Omega$ will be the 
appropriate quantity to consider. One notes that the contribution 
to the average of those addenda of ${\bf T}^m_{k;m}$ that can be written 
as an overall ordinary derivative with respect to some coordinate 
will be negligible, since $s^{ik}$ and $a^{ik}$ have the assumed periodic 
behaviour at a microscopic scale. By recalling the definitions (9) 
and (14) one can bring the conservation identity (15) to the form

$$16\pi{\bf T}^m_{k;m}=2({\bf g}^{[mi]}B_{[ik]})_{,m}
+({\bf g}^{[im]}B_{[im]})_{,k}
-{\bf g}^{[im]}_{~~~,k}B_{[im]};\eqno(40)$$
hence, whenever the globally differentiated terms provide a 
negligible contribution to the average, one can write
$$<{\bf T}^m_{k;m}>
=-(1/{16\pi})<{\bf g}^{[im]}_{~~~,k}B_{[im]}>,\eqno(41)$$
which immediately reveals that, under the above mentioned 
conditions, the $k$-th component of the mean generalized force 
density vanishes if ${\bf g}^{[im]}$ does not depend on the $k$-th 
coordinate.\par
     Let us consider a quantity $Q$, expressed in terms of the $s^{ik}$ 
and of the $a^{ik}$ defined by (37), and homogeneous with respect to 
differentiation, like all the geometric objects that we are 
considering. If $Q$ is differentiated once with respect to $x^m$, 
the resulting quantity will be the sum of terms each containing a 
number of factors $k^A_m$ increased by one with respect to the number 
of such factors appearing in $Q$. We indicate generically with 
$[b]$ a quantity of the same order of magnitude as $b^{ik}_A$ or 
$c^{ik}_A$, and 
with $[k]$ a quantity having the same order of magnitude as $k^A_m$. 
Then the largest term in the first derivatives of $s^{ik}$, of $a^{ik}$, 
of $s_{ik}$, of $a_{ik}=s_{ip}s_{kq}a^{pq}$ and of ${\bf g}^{ik}$ with respect to $x^m$ is a quantity 
whose magnitude can be indicated with $[kb]$.\par
     The term within brackets at the right-hand side of (41) is 
homogeneous of degree three with respect to differentiation; it 
will contain leading terms of the type $[k^3b^2]$, smaller terms like 
$[k^3b^3]$ and so on with higher powers of $[b]$. We decide for now to 
stop the calculation at the terms of magnitude $[k^3b^3]$, hence we 
need to know

$$B_{[ik]}=\Gamma^a_{[ik],a}
-\Gamma^a_{[ib]}\Gamma^b_{(ak)}-\Gamma^a_{(ib)}\Gamma^b_{[ak]}
+\Gamma^a_{[ik]}\Gamma^b_{(ab)}\eqno(42)$$
up to terms $[k^2b^2]$. The affine connection $\Gamma^i_{km}$ is defined by 
(10); since the largest term in ${\bf g}^{qr}_{~~,p}$ is a quantity of type 
$[kb]$, while the largest term in ${\bf g}^{ik}$ is a quantity of the 
order unity, in general the largest term of $\Gamma^i_{km}$ will be of 
the type $[kb]$. Therefore, in order to evaluate up to $[k^2b^2]$ all 
the terms at the right-hand side of (42), one needs to know the 
$\Gamma^i_{(km)}$ and the $\Gamma^i_{[km]}$ appearing in the products up to 
quantities $[kb]$; $\Gamma^i_{[km]}$ in the differentiated term $\Gamma^a_{[ik],a}$ is instead required up to quantities 
of type $[kb^2]$.\par
When terms up to $[kb]$ are retained, $\Gamma^i_{(km)}$ is given by the 
Christoffel symbol $\Sigma^i_{km}$ of (16), where one can replace $s^{ik}$ 
with $\eta^{ik}$ and $s_{ik}$ with the approximate form (38). In order to 
determine $\Gamma^i_{[km]}$ with the required approximation, one considers 
those equations of (10) that are skew in the upper indices:

$${\bf a}^{qr}_{~~,p}+{\bf a}^{sr}\Gamma^q_{(sp)}
+{\bf a}^{qs}\Gamma^r_{(ps)}
-{\bf a}^{qr}\Gamma^t_{(pt)}
+{\bf s}^{sr}\Gamma^q_{[sp]}
+{\bf s}^{qs}\Gamma^r_{[ps]}
=(4\pi/3)({\bf j}^q\delta^r_p-{\bf j}^r\delta^q_p).\eqno(43)$$
Since the largest term in ${\bf a}^{ik}$ is of type $[b]$, while the 
largest term in ${\bf s}^{ik}$ is of order unity, we can solve (43) for 
$\Gamma^i_{[km]}$ up to terms $[kb^2]$ if we substitute the $\Gamma^i_{(km)}$ 
appearing in it with the Christoffel symbols $\Sigma^i_{km}$ defined up 
to $[kb]$. If the exact $\Sigma^i_{km}$ are instead substituted, we find 
through exact manipulations

$$\Gamma^i_{[km]}=(1/2)(a^{~~i}_{k~;m}-a^{~~i}_{m~;k}+a^{~~~;i}_{km})
+(4\pi/3)(\delta^i_k j_m-\delta^i_m j_k).\eqno(44)$$
When $\Gamma^i_{(km)}$ is replaced in (42) by $\Sigma^i_{km}$, as it is allowed, one can write

$$B_{[ik]}=\Gamma^a_{[ik];a},\eqno(45)$$
and due to (44) $B_{[ik]}$ acquires also in the present case the 
approximate form (19). This expression for $B_{[ik]}$ contains all the 
needed terms, and also negligible ones, that will be eventually 
discarded; in the Appendix it is given explicitly as a function of 
$s^{ik}$ and of $a^{ik}$. The terms of magnitude like $[k^2b]$ occurring in $B_{[ik]}$ have the overall expression

$$(1/6)(\eta_{ip}a^{ps}_{~~,s,k}-\eta_{kp}a^{ps}_{~~,s,i})
+(1/2)\eta_{ip}\eta_{kn}\eta^{aq}a^{pn}_{~~,a,q};\eqno(46)$$
they all vary with the coordinates through a ``sin'' dependence. As 
regards their trigonometric behaviour, the individual addenda of 
$B_{[ik]}$ whose magnitude is like $[k^2b^2]$ can instead be grouped 
in two categories. To the first category, with ``sin.sin'' dependence, 
either belong terms displaying the product of a second derivative 
of $s_{mn}$ times $a^{pq}$, or terms containing the product of a second 
derivative of $a^{mn}$ times the part of magnitude $[b]$ of $s_{pq}$, defined by (38); the second category contains the remaining addenda, with 
``cos.cos'' dependence, that display the product of a first 
derivative of $s_{mn}$ times a first derivative of $a^{pq}$.\par
     In the term within brackets at the right-hand side of (41) 
the skew tensor $B_{[im]}$, whose trigonometric behaviour has been just 
examined, is contracted with ${\bf g}^{[im]}_{~~~~,k}$, which contains terms like $[kb]$, that display ``cos'' dependence, and terms of magnitude 
$[kb^2]$, which have ``sin.cos'' dependence, since they are either a 
product of $a^{im}$ times the first derivative of $s_{pq}$, or the product 
of $a^{im}_{~~,k}$ times the $[b]$ part of $s_{pq}$. Therefore the contraction 
${\bf g}^{[im]}_{~~~,k}B_{[im]}$ will contain terms like $[k^3b^2]$, arising from the 
product of a first and a second derivative of $a^{pq}$, hence 
displaying a ``sin.cos'' dependence, and terms of type $[k^3b^3]$, in 
which it will appear either the product of two sines times a 
cosine, or the product of three cosines.\par
     Certain conclusions about the mean generalized force density can 
be drawn without explicitly calculating the right-hand side of 
(41). The overall expression for the terms of magnitude $[k^3b^2]$ 
occurring in ${\bf g}^{[im]}_{~~~,k}B_{[im]}$ is\par

$$a^{im}_{~~,k}[(1/6)(\eta_{ip}a^{ps}_{~~,s,m}-\eta_{mp}a^{ps}_{~~,s,i})
+(1/2)\eta_{ip}\eta_{nm}\eta^{aq}a^{pn}_{~~,a,q}];\eqno(47)$$
after a rearrangement that puts globally differentiated terms in 
evidence, this expression takes the form

$$(1/3)(a^{~m}_{p~~,k}a^{ps}_{~~,s})_{,m}
-(8\pi^2/3)(j_pj^p)_{,k}
+(1/2)\eta^{aq}a_{pn,k}a^{pn}_{~~,a,q},\eqno(48)$$
where the indices are moved with $\eta_{ik}$. Since also the last term 
in (48) can be turned into a sum of globally differentiated 
quantities, one concludes that the terms of magnitude $[k^3b^2]$ do 
not contribute to $<{\bf T}^m_{k;m}>$; we note that in these terms only 
waves of $a^{ik}$ can appear, i.e. the wavy behaviour of the metric has 
no r\^ole in them.\par
     The terms of magnitude $[k^3b^3]$ occurring in 
${\bf g}^{[im]}_{~~~,k}B_{[im]}$ depend on the coordinates only through the trigonometric factors

$$\eqalign{cos(k^A_p x^p+\varphi^A)sin(k^B_q x^q+\varphi^B)
sin(k^C_r x^r+\varphi^C),\cr
cos(k^A_p x^p+\varphi^A)cos(k^B_q x^q+\varphi^B)
cos(k^C_r x^r+\varphi^C),\cr }
\eqno \hbox{$(49)$} $$
as already observed; a nonzero contribution to $<{\bf T}^m_{k;m}>$ coming 
from these terms can only take place if the averages of the 
trigonometric factors (49) are not all vanishing. From (A6) and 
(A7) of the Appendix one recognizes that the latter averages will 
vanish unless, for some choice of $A$, $B$, $C$, one of the 
following occurrences is realized:

$$k^A_m\pm k^B_m\pm k^C_m=0~~~
and~~~\varphi^A\pm\varphi^B\pm\varphi^C\neq(n+1/2)\pi\eqno(50)$$
for all the values of $m$, and with integer $n$; it is intended 
that the signs in front of a phase $\varphi^E$ and in front of the 
corresponding wave four-vector $k^E_m$ are always chosen in the same 
way. One concludes that the terms of magnitude $[k^3b^3]$ at the 
right-hand side of (15) cannot produce a net average generalized 
force density unless the three-waves resonance condition (50) is 
satisfied for some choice of $A$, $B$ and $C$; in this case one 
and just one of the three waves is necessarily contributed by the 
metric $s_{ik}$; the sign and the value of the individual trigonometric 
term will be decided by the combination of the phases of the three 
waves whose wave four-vectors fulfil the resonance condition. By 
specializing (37) to a particularly simple instance one can 
ascertain that $<{\bf T}^m_{k;m}>$ can actually be nonvanishing; we reach 
the conclusion that the geometric objects of Einstein's unified 
field theory may be used to represent the production of a 
macroscopic generalized force density in a given spacetime region 
through the resonance occurring at a microscopic scale between two 
progressive waves of $a^{ik}$ and one progressive wave of $s^{ik}$. From the way kept in achieving this result one expects that the production 
of net generalized forces through resonant processes in which more 
than three waves are involved can be demonstrated if one pushes to 
higher order in $[b]$ the approximation with which ${\bf g}^{ik}$ and 
$B_{[ik]}$ are calculated.\par\vbox to 1.2 cm{}

\noindent{\gr 7. A PARTICULAR EXAMPLE OF RESONANT POWER ABSORPTION OR 
EMISSION}\par\vbox to 0.8 cm {}

The resonance condition $k^A_m\pm k^B_m\pm k^C_m=0$ is just the four-
dimensional expression of the quantum mechanical rule known in the 
particular case of the frequencies as Bohr's condition. According 
to quantum physics this resonance condition plays a fundamental 
r\^ole for the exchanges of energy and momentum going on within 
matter; despite the utter differences in the variables involved 
and in the physical interpretation, the same condition is found 
necessary for the appearance of a nonvanishing $<{\bf T}^m_{k;m}>$, when 
${\bf g}^{ik}$ is endowed with the wavy microscopic structure 
prescribed by (37). We still need to prove that the generalized forces associated 
to the three-waves resonances can be really nonvanishing; we shall 
do so through a particular example, freely sketched after the 
theoretical model that wave mechanics provides for the elementary 
processes of power absorption and emission in matter: an ``atom'' at 
rest is supposed to execute simultaneously two normal vibrations 
whose angular frequencies $\omega_1$ and $\omega_2$, in keeping with 
the relativistic description, are very large when compared to the 
angular frequency $\omega$ of a ``light wave'' that interacts with 
the atomic system. We lack at present field equations for ${\bf g}^{ik}$ that could describe this microscopic occurrence in a 
consistent way; we shall limit ourselves to render some of its 
tracts through the geometric objects of Einstein's unified field 
theory as follows.\par
     With respect to the system of coordinates $x^i$, the metric 
$s^{ik}$ is assumed to display very small deviations from the Minkowski 
form; its dependence on the coordinates is given by

$$s^{ik}=\eta^{ik}
+u^{ik}_A(x^\mu)sin(\omega^A t+\varphi^A(ik));\eqno(51)$$
Greek letters again indicate the spatial coordinates, while 
$t\equiv x^4$ stands for the time coordinate, and $A=1,2$ labels 
the normal vibrations. The components of $u^{ik}_A=u^{ki}_A$ and their 
derivatives are assumed to vanish everywhere in spacetime, except 
within a world tube $\Pi$, whose spatial section $\Sigma$ at 
$x^4=$const. is compact and of atomic size; there the $u^{ik}_A$ are so 
small that can be dealt with as first order infinitesimals. The 
positive angular frequencies $\omega_A$ are nearly equal, and we 
choose $\omega_1\geq\omega_2$, while $\varphi^A(ik)=\varphi^A(ki)$ 
represent constant phases, that can take different values in 
different components of $u^{ik}_A$. We assume further that 
$a^{ik}$ can be written as

$$a^{ik}=b^{ik}+c^{ik},\eqno(52)$$
where

$$b^{ik}=v^{ik}_A(x^\mu)sin(\omega^A t+\chi^A(ik)),\eqno(53)$$
and

$$c^{ik}=d^{ik}sin(k_mx^m+\psi(ik)).\eqno(54)$$
The components of $v^{ik}_A=-v^{ki}_A$ and their derivatives are 
everywhere vanishing, except within the world tube $\Pi$, where the 
$v^{ik}_A$ can be treated as first order infinitesimals; $\chi^A(ik)=\chi^A(ki)$ are 
constant phases, that can be different for different components of 
$v^{ik}_A$; the normal vibrations of $b^{ik}$ occur with the angular 
frequencies $\omega_1$, $\omega_2$ exhibited also by the metric 
$s^{ik}$. In (54), the components of $d^{ik}=-d^{ki}$ are small constants that can be 
considered as first order infinitesimals, while $k_m$ is a 
four-vector, null with respect to the average metric 
$\bar s_{ik}=\eta_{ik}$, and 
$\psi(ik)=\psi(ki)$ are constant phases; therefore $c^{ik}$ can behave 
as the components of {\bf D} and {\bf H} do, according to Maxwell's theory, 
for an electromagnetic plane wave in vacuo; furthermore, $k_m$ is so 
chosen that the wavelength of $c^{ik}$ is large with respect to the 
spatial extension of $\Sigma$ and the positive angular frequency 
$k_4\equiv\omega$ is very small with respect to both $\omega_1$ and 
$\omega_2$, its order of magnitude being the same as for the 
difference $\omega_1-\omega_2$. We assume eventually that in our 
units $\vert s^{ik}_{~~,\mu}\vert$ and $\vert b^{ik}_{~~,\mu}\vert$ are small, when compared 
to $\vert s^{ik}_{~~,4}\vert$ and to $\vert b^{ik}_{~~,4}\vert$, since in the relativistic 
wavefunction of an atom the characteristic length for the spatial 
dependence is the Bohr radius, while the scale of the time 
dependence is provided by the Compton period.\par
The power absorption or emission by the ``atom'' will be detected 
through the average $<{\bf T}^m_{4;m}>$ extended to a box $\Omega$ which 
encloses the world tube $\Pi$ for a span of the time coordinate 
that is very long with respect to the period $T=2\pi/\omega$ of 
$c^{ik}$ ; in these conditions the contribution to the average coming 
from the terms that are globally differentiated can be 
disregarded. Then we can avail of (41) and write

$$<{\bf T}^m_{4;m}>=-(1/16\pi)<[(-s)^{1/2}_{~~~,4}(b^{im}+c^{im})
+(-s)^{1/2}(b^{im}+c^{im})_{,4}]B_{[im]}>.\eqno(55)$$
This can hardly be called a macroscopic average, but one can 
readily imagine the extension of the present argument to an 
assembly of independent ``atoms''. A calculation adequate to reveal 
a resonant absorption or emission of power can be done through the 
approximation scheme of the previous Section; again, resonant 
processes in which only two oscillations take part are ruled out, 
and the next available possibility is a resonance in which three 
oscillations are involved, of which one and just one belongs to 
the metric.\par
Due to the choices done for the time dependence of $s^{ik}$ and of $a^{ik}$, 
a resonant three-waves absorption or emission of power can only 
occur if $\omega=\omega_1-\omega_2$, and the only terms in 
$<{\bf T}^m_{4;m}>$ of relevance in this process will be those that are written as a triple product, in which one factor is provided by $s_{ik}$, a second one by $b^{ik}$, and a third one by $c^{ik}$; of these terms, the ones that will contribute with the greatest strength will be those whose 
generic form reads

$$s_{ab,4,4}b^{cd}_{~~,4}c^{ef}~~~
or~~~s_{ab,4}b^{cd}_{~~,4,4}c^{ef},\eqno(56)$$
i.e. those terms in which $s_{ik}$ and $b^{ik}$, whose dependence on time 
was assumed to be faster than the spatial one, and also much 
faster than the spacetime dependence of $c^{ik}$, are collectively 
differentiated as many times as possible with respect to $x^4$.\par
The first and the last addendum at the right-hand side of (55) 
cannot produce terms with the forms (56) and can be dropped if one 
wishes to retain only the largest contributions, as we shall do. 
By availing of (A5) one finds the approximate 
expression

$$\eqalign{<{\bf T}^m_{4;m}>&=-(1/16\pi)<\eta^{pq}s_{pq,4}
[(1/4)c^{\lambda\mu}b_{\lambda\mu,4,4}
+(2/3)c^{4\mu}b_{4\mu,4,4}]\cr
&+b^{4\mu}_{~~,4}[c^4_{~\mu}s_{44,4,4}+c^{4\rho}s_{\mu\rho,4,4}
-(1/3)c^4_{~\mu}\eta^{pq}s_{pq,4,4}]\cr
&+(1/2)b^{\lambda\mu}_{~~,4}[c^4_{~\mu}s_{\lambda 4,4,4}
-c^4_{~\lambda}s_{\mu 4,4,4}]>, }
\eqno \hbox{$(57)$} $$
where indices have been lowered with $\eta_{ik}$. A specialization of 
this result that may be of some interest is attained if we assume 
that $s^{ik}$, whose form is given by (51), is conformally flat, i.e. 
if we can write also $s^{ik}=e^{-\sigma}\eta^{ik}$, with 
$\vert \sigma\vert \ll 1$. 
Then $<{\bf T}^m_{4;m}>$, after neglecting globally differentiated terms 
and by retaining only the largest contributions, gets the simple 
expression:

$$<{\bf T}^m_{4;m}>=-(1/16\pi)<\sigma_{,4}c^{ik}b_{ik,4,4}>,\eqno(58)$$
from which it is recognized that a nonvanishing average absorption 
or emission of power can indeed take place, provided that the 
resonance condition $\omega=\omega_1-\omega_2$ is satisfied.\par
     The essential r\^ole played by the microscopic behaviour of the 
metric elicits a comment of a general character. If we consider 
the average generalized force produced by some fields defined on a 
rigid Minkowski background, we come up with an expression like 
$<T^m_{k,m}>$, where $T^m_{k,m}$ is the ordinary divergence of some 
energy tensor 
$T^m_k$; then, since $\int T^m_{k,m}d\Omega$ can be transformed into a surface integral over the boundary $\Delta$ of the domain of integration 
$\Omega$, the detailed behaviour inside $\Omega$, in particular a 
resonant behaviour of the fields that enter the definition of $T^m_k$, 
is completely irrelevant to the average: provided that the values 
of the fields and of their derivatives appearing in $T^m_k$ were kept 
unaltered on $\Delta$, the value of $<T^m_{k,m}>$ would remain the same 
also if the resonant behaviour of the fields were substituted with 
an incoherent one. The appearance of the covariant divergence 
${\bf T}^m_{k;m}$ in the differential conservation laws of the general 
relativistic theories, as it occurs in (15), can make the 
difference: if ${\bf T}^m_{k;m}$ cannot be transformed into a sum of 
globally differentiated addenda, a direct relation becomes 
possible between a resonant microscopic behaviour of the fields, 
inclusive of the metric, and the average generalized 
force.\vfill\break

\noindent{\gr 8. CONCLUDING REMARKS}\par\vbox to 0.8 cm {}

     Through particular examples obtained by imposing a priori some 
behaviour on the geometric objects of Einstein's unified field 
theory it has been shown how deep an influence a microscopic 
structure of the metric can exert on the macroscopic appearances 
that constitute the world of experience. Fluctuations of the 
metric with a very small amplitude and with a microscopic 
characteristic length are in fact capable of ruling the 
constitutive relation of macroscopic electromagnetism in 
nonconducting, nondispersive media; moreover, a microscopic wavy 
behaviour of the metric and of the field ${\bf a}^{ik}$ can result in 
the production of macroscopic generalized forces through 
three-waves resonance processes in which the wavevectors and the 
frequencies involved obey the very conditions that, according to 
quantum physics, rule the exchanges of energy and momentum 
occurring within matter.\par
     It appears that the geometric objects of Einstein's unified field 
theory indeed offer entirely new opportunities for describing the 
macroscopic reality by starting from hypothetic microscopic 
structures and processes. The heuristic method adopted in the 
present paper is however of very limited scope: a priori 
assumptions for ${\bf g}^{ik}$ may suggest interesting possibilities, 
but in order to proceed further, field equations dictating the 
spacetime behaviour of ${\bf g}^{ik}$ need to be assigned and solved. 
As previously mentioned, such equations are presently lacking: 
unfortunately, we cannot rely on the ones proposed by Einstein 
[10] since, with the interpretation of the geometric objects 
proposed here, those equations imply that $T_{ik}$, ${\bf j}^i$ and $K_{ikm}$ are vanishing everywhere. Only through field equations one can 
hope to develop a theory in which the outcomes of the previous 
Sections would be properly framed. Writing down sensible equations 
is of course a quite difficult task, but possibly not a desperate 
one: the whole wealth of experimental information of atomic and 
condensed matter physics is at our disposal as a guide in this 
endeavour.\vfill\break

\noindent{\gr APPENDIX}\par\vbox to 0.8 cm {}

When terms up to $[kb^2]$ are retained, one writes

$$4\pi j_i=a^{~s}_{i~;s}=s_{ip}a^{ps}_{~~,s}
+(1/2)\eta_{ip}\eta^{sr}a^{pn}s_{sr,n};\eqno(A1)$$
up to terms of magnitude $[k^2b^2]$, one finds

$$a^{~n}_iS_{nk}=(1/2)\eta_{ip}\eta^{sr}a^{pn}
(s_{nr,k,s}+s_{kr,n,s}-s_{nk,r,s}-s_{sr,n,k}),\eqno(A2)$$

$$a^{pq}S_{pqik}=a^{rn}(s_{ir,n,k}-s_{kr,n,i}),\eqno(A3)$$
and

$$\eqalign{a^{~~;a}_{ik;a}&=s_{ip}s_{kn}s^{aq}a^{pn}_{~~,a,q}\cr
&+\eta^{aq}a^{rn}_{~~,q}[\eta_{kn}(s_{ri,a}+s_{ai,r}-s_{ra,i})
-\eta_{in}(s_{rk,a}+s_{ak,r}-s_{ra,k})]\cr
&-(1/2)\eta_{ip}\eta_{kn}\eta^{aq}\eta^{rs}a^{pn}_{~~,r}
(2s_{as,q}-s_{aq,s})\cr
&+(1/2)\eta^{aq}a^{rn}[\eta_{kn}(s_{ri,a}+s_{ai,r}-s_{ra,i})
-\eta_{in}(s_{rk,a}+s_{ak,r}-s_{ra,k})]_{,q}.\cr }
\eqno \hbox{$(A4)$} $$

Hence the sought for expression of $B_{[ik]}$ reads:

$$\eqalign{B_{[ik]}&=(1/6)[(s_{ip,k}-s_{kp,i})a^{ps}_{~~,s}
+s_{ip}a^{ps}_{~~,s,k}-s_{kp}a^{ps}_{~~,s,i}]\cr
&+(1/12)\eta^{sr}[\eta_{ip}(a^{pn}s_{sr,n})_{,k}
-\eta_{kp}(a^{pn}s_{sr,n})_{,i}]\cr
&+(1/2){\big \{}(1/2)\eta^{sr}a^{pn}[\eta_{ip}(2s_{kr,n,s}-s_{sr,n,k})
-\eta_{kp}(2s_{ir,n,s}-s_{sr,n,i})]\cr
&+a^{rn}(s_{ir,n,k}-s_{kr,n,i})+s_{ip}s_{kn}s^{aq}a^{pn}_{~~,a,q}\cr
&+\eta^{aq}a^{rn}_{~~,q}[\eta_{kn}(s_{ri,a}+s_{ai,r}-s_{ra,i})
-\eta_{in}(s_{rk,a}+s_{ak,r}-s_{ra,k})]\cr
&-(1/2)\eta_{ip}\eta_{kn}\eta^{aq}\eta^{rs}a^{pn}_{~~,r}
(2s_{as,q}-s_{aq,s}){\big \}}.\cr }
\eqno \hbox{$(A5)$} $$

Two trigonometric relations are recalled for convenience:

$$\eqalign{cos\alpha sin\beta sin\gamma=
-(1/4)&[cos(-\alpha+\beta+\gamma)-cos(\alpha-\beta+\gamma)\cr
&-cos(\alpha+\beta-\gamma)+cos(\alpha+\beta+\gamma)],\cr }
\eqno \hbox{$(A6)$} $$

$$\eqalign{cos\alpha cos\beta cos\gamma=
(1/4)&[cos(-\alpha+\beta+\gamma)+cos(\alpha-\beta+\gamma)\cr
&+cos(\alpha+\beta-\gamma)+cos(\alpha+\beta+\gamma)].\cr }
\eqno \hbox{$(A7)$} $$
\vfill\break

\noindent{\gr REFERENCES}\par\vbox to 0.4 cm {}

\noindent [1]~Lorentz, H.A. (1902). {\sl Proc. Roy. Acad. Amsterdam}, 254; {\sl id.} (1952). {\sl The Theory of Electrons} (Dover Publications, 
New York).\par\noindent
[2]~Einstein, A. (1919). {\sl S. B. Preuss. Akad. Wiss.} {\bf 20}, 
349; {\sl id.} (1923). {\sl S. B. Preuss. Akad. Wiss.} {\bf 33}, 
359.\par\noindent
[3]~Klein, O. (1926). {\sl Zeitschr. f. Phys.} {\bf 37}, 
895.\par\noindent
[4]~Schouten, J.A. (1954). {\sl Ricci-Calculus} (Springer-Verlag, 
Berlin), p. 67.\par\noindent
[5]~Lanczos, C. (1942). {\sl Phys. Rev.} {\bf 61}, 713; {\sl id.} 
(1957). {\sl Rev. Mod. Phys.} {\bf 29}, 337.\par\noindent
[6]~Lanczos, C. (1963). {\sl J. Math. Phys.} {\bf 4}, 951; {\sl id.} 
(1966). {\sl J. Math. Phys.} {\bf 7}, 316.\par\noindent
[7]~Nelson, A. H. (1972). {\sl Mon. Not. R. astr. Soc.} {\bf 158}, 
159; Marochnik, L.S., Pelichov, N.V., and Vereshkov, G.M. (1975). 
{\sl Astrophys. Space Sci.} {\bf 34}, 249, 281; Ellis, G.F.R. 
(1984). In {\sl Proceedings of the Tenth International Conference 
on General Relativity and Gravitation}, Bertotti, B., de Felice, 
F., and Pascolini, A., eds. (Reidel, Dordrecht); Carfora, M., and 
Marzuoli, A. (1984). { Phys. Rev. Lett.} {\bf 53}, 2445; Schmidt, 
H.-J. (1988). { Astron. Nachr.} {\bf 309}, 307.\par\noindent
[8]~Zalaletdinov, R.M. (1992). {\sl Gen. Rel. Grav.} {\bf 24}, 
1015.\par\noindent
[9]~Noonan, T.W. (1984). {\sl Gen. Rel. Grav.} {\bf 16}, 
1103.\par\noindent
[10]~Einstein, A. (1925). {\sl S. B. Preuss. Akad. Wiss.} {\bf 22}, 
414; Einstein, A., and Straus, E.G. (1946). {\sl Ann. Math.} {\bf 
47}, 731; Einstein, A. (1948). {\sl Rev. Mod. Phys.} {\bf 20}, 35; 
Einstein, A., and Kaufman, B. (1955). {\sl Ann. Math.} {\bf 62}, 
128.\par\noindent
[11]~Schr\"odinger, E. (1947). {\sl Proc. R. I. Acad.} {\bf 51A}, 
163, 205; {\sl id.} (1948). {\sl Proc. R. I. Acad.} {\bf 52A}, 
1.\par\noindent
[12]~H\'ely, J. (1954). {\sl Comptes Rend. Acad. Sci. (Paris)} {\bf 
239}, 747.\par\noindent
[13]~Schr\"odinger, E. (1954). {\sl Space-Time Structure} 
(Cambridge University Press, Cambridge), p. 99.\par\noindent
[14]~Synge, J.L. (1966). {\sl What is Einstein's Theory of 
Gravitation?}, from: {\sl Perspectives in Geometry and Relativity 
(Essays in Honor of V. Hlavat\'y)} (Indiana University Press, 
Bloomington), p. 7.\par\noindent
[15]~Antoci, S. (1991). {\sl Gen. Rel. Grav.} {\bf 23}, 
47.\par\noindent
[16]~Kur\c suno\u glu, B. (1952). {\sl Phys. Rev.} {\bf 88}, 1369; H\'ely, 
J. (1954).  {\sl Comptes Rend. Acad. Sci. (Paris)} {\bf 239}, 
385.\par\noindent
[17]~Tonnelat, M.A. (1955). {\sl La th\'eorie du champ unifi\'e  
d'Einstein} (Gauthier-Villars, Paris), p. 37.\par\noindent
[18]~Hlavat\'y, V. (1957). {\sl Geometry of Einstein's Unified Field  
Theory} (Noordhoff, Groningen), Chap. II.\par\noindent
[19]~Borchsenius, K. (1978). {\sl Nuovo Cimento} {\bf 46A}, 
403.\par\noindent
[20]~Einstein, A. (1955). {\sl The Meaning of Relativity} 
(Princeton  University Press, Princeton), p. 149.\par\noindent
[21]~Synge, J.L. (1960). {\sl Relativity: The General Theory} 
(North-Holland, Amsterdam), p. 317.\par\noindent
[22]~Antoci, S. (1992). {\sl Prog. Theor. Phys.} {\bf 87}, 
1343.\par\noindent
[23]~Post, E.J. (1962). {\sl Formal Structure of Electromagnetics} 
(North-Holland, Amsterdam), Chap. VI.\end